\documentclass[fleqn,10pt]{wlscirep}

\usepackage{float}
\usepackage{times}
\usepackage{graphicx}
\usepackage{amsfonts}
\usepackage{amsmath, amsthm, amssymb}
\usepackage{dsfont}
\usepackage{mathptm}
\usepackage{color}
\usepackage{subcaption}
\usepackage{tikz}

\newcommand{\eg}{\textit{e.g. }}
\newcommand{\etal}{\emph{et al.}}
\def\i{\mathrm{i}}

\newcommand{\g}{\underline{\gamma}}
\newcommand{\gt}{\underline{\tilde{\gamma}}}
\newcommand{\N}{\underline{\mathcal{N}}}
\newcommand{\Nt}{\underline{\tilde{\mathcal{N}}}}

\title{General solution of 2D and 3D superconducting quasiclassical systems: coalescing vortices and nanoisland geometries}

\author[1]{Morten Amundsen}
\author[1]{Jacob Linder}
\affil[1]{Department of Physics, Norwegian University of
Science and Technology, N-7491 Trondheim, Norway. Correspondence should be addressed to J. L. (email: jacob.linder@ntnu.no).}

\begin{abstract}
An extension of quasiclassical Keldysh-Usadel theory to higher spatial dimensions than one is crucial in order to describe physical phenomena like charge/spin Hall effects and topological excitations like vortices and skyrmions, none of which are captured in one-dimensional models. We here present a numerical finite element method which solves the non-linearized 2D and 3D quasiclassical Usadel equation relevant for the diffusive regime. We show the application of this on three model systems with non-trivial geometries: (i) a bottlenecked Josephson junction with external flux, (ii) a nanodisk ferromagnet deposited on top of a superconductor and (iii) superconducting islands in contact with a ferromagnet. In case (i), we demonstrate that one may control externally not only the geometrical array in which superconducting vortices arrange themselves, but also to cause coalescence and tune the number of vortices. In case (iii), we show that the supercurrent path can be tailored by incorporating magnetic elements in planar Josephson junctions which also lead to a strong modulation of the density of states. The finite element method presented herein paves the way for gaining insight in physical phenomena which have remained largely unexplored due to the complexity of solving the full quasiclassical equations in higher dimensions.
\end{abstract}
\begin{document}

\flushbottom
\maketitle

\thispagestyle{empty}

\section*{Introduction}

Nonlinear differential equations (NLDEs) play a pivotal role in virtually all areas of physics. They are used to describe completely disparate phenomena ranging from the behavior of ocean waves to the elasticity of materials. Thus, techniques to solve such equations are of general interest as they provide a way to obtain insight in a number of different physical systems. NLDEs are known for being notoriously difficult to solve and, more often than not, a set of NLDEs describing a particular physical scenario has to be addressed as a distinct problem since general techniques to solve such equations are scarce. 

In quantum condensed matter physics, mesoscopic systems both in and out of equilibrium represent a very important arena where NLDEs are prevalent. A powerful tool used to describe such systems is the quasiclassical Keldysh theory, which has been reviewed in several works \cite{schmid_nato_81, serene_physrep_83, larkin_elsevier_86, rammer_rmp_86, sauls_94, belzig_superlattice_99, kopnin_book_09}. The theory is based on a Green function method which thus has a natural way of including disorder and other types of self-energies in the system. The quasiclassical Keldysh theory is capable of treating both ballistic systems and "dirty" systems. In the latter case, quasiparticles are elastically scattered within the mean free path $l_\text{mfp}$ causing the resulting motion to be diffusive. In essence, the quasiclassical theory is a perturbation expansion valid when all energy scales in the problem are much smaller than the Fermi energy $E_F$. Conversely, all length scales in the system should be much larger than the Fermi wavelength. This situation is realized in a number of mesoscopic systems, including normal metals, superconductors and weakly polarized ferromagnets. Strongly polarized ferromagnets, where the exchange energy splitting $h$ of the majority and minority spin bands is comparable in size to the Fermi energy $E_F$, appear to be at odds with the applicability of quasiclassical theory. However, there also exists a way in which such systems can be described in this framework. When the splitting $h$ is sufficiently large, the two spin-bands may be treated separately so that $h$ does not enter the problem at all and one is left with two decoupled spin species \cite{eschrig_prl_09}. Such an approach is also suitable to address extreme cases such as half-metals using quasiclassical theory, as done recently in Refs. \cite{eschrig_njp_15, buzdin_prb_15}.

The equation of motion for the central object in quasiclassical Keldysh theory, the Green function $\check{g}$, is a NLDE (known as the Eilenberger \cite{eilenberger} equation for arbitrary impurity scattering and the Usadel \cite{usadel}  equation in the diffusive limit) and must be supplemented by appropriate boundary conditions. Focusing on the diffusive limit, as it is often the experimentally relevant one, a variety of options are available depending on the physical situation at hand. In the simplest case of perfectly transparent interfaces, the Green function is taken as continuous across the interface. This is clearly an idealized scenario and the more realistic Kupriyanov-Lukichev \cite{kl} boundary conditions describe an interface in the tunneling limit where there exists a substantial interface resistance. Boundary conditions for an arbitrary interface transparency were developed in Ref. \cite{nazarov}. When the interface has magnetic properties, either because of an intrinsically thin magnetic layer inserted between \eg two metals or superconductors or if one of the regions separated by the interface is magnetic on its own, one must use spin-dependent boundary conditions. Pioneered in Refs. \cite{mills_prb_88, tokuyasu_prb_88}, these were brought to a more tractable form by Cottet and co-workers in the diffusive limit \cite{cottet_prb_09}. However, up until recently there existed a knowledge gap in terms of how to describe strongly polarized magnetic interfaces in quasiclassical theory. Eschrig \etal~solved this problem in Ref. \cite{eschrig_njp_15}.

It is clear that the development of a numerical routine that is able to solve the quasiclassical Keldysh equations in higher dimensions than 1D will be of great value in terms of studying a vast number of physical phenomena, including various types of Hall effects, spin swapping, and topological excitations such as magnetic skyrmions and vortices. None of these phenomena can be captured in an effective 1D model. Furthermore, the ability to handle complex higher dimensional geometries numerically allows for the modeling of systems which are more closely related to experiments. For instance, superconducting nanoisland systems and vortices in mesoscopic structures have received much attention experimentally\cite{cren_prl_09, kim_np_2012, serrier-garcia_prl_2013,cherkez_prx_2014, stepniak_aipa_2015}. These systems require not only solution in 2D or 3D, but also the description of non-trivial geometries within the numerical framework. Such solutions have been investigated using the Ginzburg-Landau formalism in the context of flux patterns and vortex states in superconductors \cite{schweigert_prb_98, schweigert_prl_98, berdiyorov_prl_09}. The ability to aid experiments with numerical routines that are both geometry and dimension independent would be highly beneficial to their study. Nevertheless, explicit solutions of the full quasiclassical equations in two dimensions have rarely been reported \cite{cuevas_prl_07, cuevas_jlt_08}. In the linearized regime, corresponding to a weak proximity effect, several works have considered the 2D solution of the Usadel equation \cite{crouzy_prb_07, bakurskiy_sst_13, volkov_prb_13, alidoust_prl_12, alidoust_15}. Motivated by this, we report as the main result of this paper the description of a finite element method that we have developed which is capable describing mesoscopic systems in 2D and 3D using quasiclassical theory without any linearization. As far as the authors are aware, this is the first work to solve the Usadel equations in 3D. After going through the details of this method, we show its application to three model systems. One of our main findings is that in a 2D Josephson junction exposed to a magnetic flux, it is possible to control not only the geometrical array in which superconducting vortices arrange themselves, but it is also possible to cause coalescence and thus tune the number of vortices. In addition, we show that the supercurrent flow through planar junction geometries can be tailored by the magnetization pattern and strength and also spatially modulates the proximity-induced density of states, which can be probed by STM-measurements. We organize our presentation as follows.  First, we introduce the system of coupled NLDEs that define the central equations in quasiclassical theory. The finite element method solving these equations in 2D and 3D is described in detail in the next section. We proceed to show the application of this method to three different hybrid structures where a superconducting material is coupled to a normal metal with external flux, and to a ferromagnet respectively. Finally, we provide a discussion of our results and concluding remarks.

 \section*{Theory}\label{sec:theory}

In this section, we write down the quasiclassical equation of motion for $\check{g}$ in the diffusive limit and its belonging boundary conditions. The task at hand is then to solve this numerically in 2D and 3D, and we demonstrate how this can be accomplished using a finite element method in the next section. 

$\check{g}$ is an $8\times8$ matrix satisfying $\check{g}^2 = \check{1}$ with the following structure:
\begin{align}\label{eq:g}
\check{g} = \begin{pmatrix}
\hat{g}^R & \hat{g}^K \\
\hat{0} & \hat{g}^A \\
\end{pmatrix},
\end{align}
where $\hat{g}^{R,A,K}$ are the retarded, advanced, and Keldysh $4\times4$ Green function matrices. The relation $\hat{g}^A = -(\hat{\rho}_3\hat{g}^R\hat{\rho}_3)^\dag$ holds both in and out of equilibrium where $\hat{\rho}_3 = \text{diag}(1,1,-1,-,1)$. The relation $\hat{g}^K = (\hat{g}^R-\hat{g}^A)\tanh(\beta\varepsilon/2)$ holds in equilibrium, so that in this scenario one only needs to determine $\hat{g}^R$ in order to completely specify $\check{g}$. The structure of the retarded Green function looks as follows:
\begin{align}
\hat{g}^R = \begin{pmatrix}
\underline{g} & \underline{f} \\
-\underline{\tilde{f}} & -\underline{\tilde{g}} \\
\end{pmatrix},\; \underline{g} = \begin{pmatrix}
g_{\uparrow\uparrow} & g_ {\uparrow\downarrow} \\
g_{\downarrow\uparrow} & g_{\downarrow\downarrow} \\
\end{pmatrix},\; \underline{f} = \begin{pmatrix}
f_{\uparrow\uparrow} & f_ {\uparrow\downarrow} \\
f_{\downarrow\uparrow} & f_{\downarrow\downarrow} \\
\end{pmatrix}.
\end{align}
where $\underline{g}=\underline{g}(\varepsilon)$ and $\underline{f}=\underline{f}(\varepsilon)$ denote the $2\times2$ normal and anomalous Green function matrices in spin space, respectively. The $\tilde{\ldots}$ operation means complex conjugation and reversal of the energy argument $\varepsilon \to (-\varepsilon)$. 

The Usadel equation reads:
\begin{align}\label{eq:usadel}
D \nabla (\check{g} \nabla \check{g}) + \i[\varepsilon\check{\rho}_3 + \check{\Sigma},\check{g}] = 0
\end{align}
where $D$ is the diffusion coefficient, $\check{\rho}_3 = \text{diag}(\hat{\rho}_3,\hat{\rho}_3)$, while $\check{\Sigma}$ is a matrix describing the self-energies of the problem. In general, it can be a functional of the Green function matrix itself, i.e. $\check{\Sigma} = \check{\Sigma}(\check{g})$. In the specific case of a ferromagnetic material, one has
\begin{align}
\check{\Sigma} = \text{diag}(\hat{M}, \hat{M}),\; \hat{M} = \vec{h}\cdot \text{diag}(\underline{\vec{\sigma}}, \underline{\vec{\sigma}}^*)
\end{align}
where $\vec{h}$ describes the magnitude and direction of the magnetic exchange field while $\underline{\vec{\sigma}} = (\underline{\sigma_x},\underline{\sigma_y},\underline{\sigma_z})$ is the vector of Pauli matrices. In the presence of gauge fields, such as a U(1) magnetic vector potential $\vec{A}$ describing an external magnetic field one has to replace the gradient operator with its covariant equivalent:
\begin{align}
\nabla \to \nabla -\i q [\vec{A}\check{\rho}_3, \check{g}]
\end{align}
where $q$ is the charge of the fermion field. A similar substitution is also made if one wishes to include an SU(2) gauge field $\vec{\underline{\mathcal{A}}}$ that describes antisymmetric spin-orbit coupling of Rashba or Dresselhaus type. In this work, we will use the standard Kupriyanov-Lukichev \cite{kl} boundary conditions as a realistic description of the interface regions. While originally derived for the tunneling regime, these boundary conditions have been shown to give good results also for moderately to highly transparent interfaces\cite{hammer_prb_2007}, which are considered herein. For an interface separating a material 1 on the left side from a material 2 on the right side, they read:
\begin{align}\label{eq:bc1}
2\zeta_j L_j \check{g}_j \vec{n}\cdot\nabla\check{g}_j = [\check{g}_1,\check{g}_2],\; j=\{1,2\}.
\end{align}
Here, $\zeta_j = R_B/R_j$ describes the ratio between the interface resistance and the bulk resistance of region $j$ while $L_j$ is the length of region $j$. Here, $\vec{n}$ is the unit vector normal to the interface pointing from region 1 to 2. At interfaces to air, no current is allowed to flow and the boundary condition is
\begin{align}\label{eq:bc2}
\vec{n}\cdot\nabla \check{g} = 0
\end{align}
where $\vec{n}$ again represents the unit vector normal to the air interface.
Equations (\ref{eq:usadel}), (\ref{eq:bc1}), and (\ref{eq:bc2}) define a system of coupled differential equations with belonging boundary conditions and the task is to find the solution $\check{g}$. For concreteness, we restrict our attention to an equilibrium scenario where only the retarded Green function matrix $\hat{g}^R$ must be found. Even with this restriction, the equations are capable of describing a variety of different mesoscopic systems. The equation system for $\hat{g}^R$ is identical to the one for $\check{g}$, as can be verified by direct insertion of Eq. (\ref{eq:g}) in the place of $\check{g}$, by replacing all $\check{\ldots}$ matrices with their $\hat{\ldots}$ equivalents. Before proceeding to a description of the finite element method we have used to solve this equation set in 2D and 3D, it is useful to introduce a Ricatti parametrization \cite{schopohl_prb_1995} of $\hat{g}^R \equiv \hat{g}$. This parametrization, first applied in Ref.\cite{konstandin_prb_05} in the context of the Usadel equation, simplifies the numerical implementation of the equations by exploiting the symmetries and normalization of $\hat{g}$. One introduces two matrices in spin-space, $\g$ and $\gt$, which define $\hat{g}$ as follows:
\begin{align}\label{eq:riccati}
\hat{g}= \begin{pmatrix}
\N(1+\g\gt) & 2\N\g \\
-2\Nt\gt & -\Nt(1+\gt\g) \\
\end{pmatrix},\; \N = (1-\g\gt)^{-1},\; \Nt = (1-\gt\g)^{-1}.
\end{align}
This parametrization satisfies both the proper symmetry relations between the elements of $\hat{g}$ as well as the normalization condition $\hat{g}^2 = \hat{1}$.

Equations (\ref{eq:usadel}), (\ref{eq:bc1}), and (\ref{eq:bc2}) comprise a set of second-order coupled partial nonlinear differential equations which, when solved, determine the Green function $\check{g}$ of the system. Various physical quantities of interest may then be computed, such as the charge current density $\vec{J}_Q$ and the density of states (DOS), given as:
\begin{align}
\vec{J}_Q &= \frac{N_0eD}{4} \int^\infty_{-\infty} d\varepsilon \text{Tr}\{\hat{\rho}_3 (\check{g}\nabla \check{g})^\text{K}\}
\end{align}

\begin{align}
\text{DOS} &= \frac{1}{2}\text{Tr}\{\N (1 + \g \gt)\}
\end{align}

Another physical quantity that may be computed is the pair correlation function, $\Psi$, indicating the degree to which superconducting correlations exist in the system. It is given as:

\begin{align}
\Psi = \frac{1}{8}\int^\infty_{-\infty} d\varepsilon \left[ \hat{g}^K(1,4) - \hat{g}^K(2,3) \right]
\end{align}
where $\hat{g}^K(i,j)$ refers to the element in column $i$ and row $j$ of the Keldysh Green function matrix.

A general analytical solution of equations (\ref{eq:usadel}), (\ref{eq:bc1}), and (\ref{eq:bc2})  is impossible. Some progress can be made by linearizing the equations, as is often done when considering a superconducting proximity effect. However, this approximation limits the validity of the obtained results and may cause the loss of novel physical phenomena that are only captured when the full equations are used. To do so, one must use a numerical approach. So far, only a handful of works have managed to solve the 2D Usadel equation numerically. This has been done in the full proximity effect regime for a superconductor/normal metal/superconductor junction in Refs. \cite{cuevas_prl_07, cuevas_jlt_08}. To the best of our knowledge, no work has ever reported a solution of the Usadel equations in 3D. 

\subsection*{Implementation of the finite element method}\label{sec:resultsnum}
We here present a way to solve the quasiclassical equations in 2D and 3D using a finite element method. Its detailed description follows below. After its presentation, we show its application to 2D and 3D model systems by solving the equations without any approximations. 

Inserting \ref{eq:riccati} into equation \ref{eq:usadel} results in the following:
\begin{equation}\label{eq:usadelriccati}
\begin{aligned}
\nabla^2\g + 2\nabla\g\cdot\Nt\gt\nabla\g - 2i\N(1 + \g\gt)\vec{A}\cdot\nabla\g -2i\vec{A}\cdot\nabla\g\Nt(1+\gt\g) - 4\vec{A}^2\g\Nt(1+\gt\g) +i\vec{h}\cdot(\underline{\vec{\sigma}}\g - \g\underline{\vec{\sigma}}^*) + 2i(\varepsilon + i\delta) \g= 0 \\
\nabla^2\gt + 2\nabla\gt\cdot\N\g\nabla\gt + 2i\Nt(1 + \gt\g)\vec{A}\cdot\nabla\gt +2i\vec{A}\cdot\nabla\gt\N(1+\g\gt) - 4\vec{A}^2\gt\N(1+\g\gt)-i\vec{h}\cdot(\underline{\vec{\sigma}}^*\gt - \gt\underline{\vec{\sigma}}) +2i(\varepsilon + i\delta) \gt = 0
\end{aligned}
\end{equation}
where $\delta$ models the effect of inelastic quasiparticle scattering (the so-called Dynes parameter \cite{dynes}). We set $\delta/\Delta = 10^{-3}$ in this paper where $\Delta$ is the bulk superconducting gap. In Eq. (\ref{eq:usadelriccati}), it is also possible to include self-energies corresponding to spin-flip and spin-orbit scattering on impurities which act pair-breaking on superconducting correlations. This typically amounts to a reduction of the magnitude of the superconducting proximity effect and we omit these terms in the present work. We also note that the effect of Rashba and Dresselhaus spin-orbit interactions were derived in Ricatti-parametrized form very recently \cite{jacobsen_prb_15a, jacobsen_prb_15b}.  As $\g$ and $\gt$ are $2\times2$ matrices, thus containing 4 elements each, it is clear that the solution of equation \ref{eq:usadelriccati} involves solving a system of 8 coupled NLDEs. For brevity, we introduce the notation

\begin{align}\label{eq:Xvec}
\chi = \begin{pmatrix}
\gamma_{11} & \gamma_{12} & \gamma_{21} & \gamma_{22} & \tilde{\gamma}_{11} & \tilde{\gamma}_{12} & \tilde{\gamma}_{21} & \tilde{\gamma}_{22} \\
\end{pmatrix}^T
\end{align}
where $\gamma_{ij}$ and $\tilde{\gamma}_{ij}$ are elements of $\g$ and $\gt$ respectively. Equation \ref{eq:usadelriccati} may then be written as

\begin{align}\label{eq:usadelreduced}
\nabla^2\chi^{(\alpha)} + F^{(\alpha)}(\g, \gt, \nabla\g, \nabla\gt) = 0
\end{align}
where $\alpha$ is an element of equation \ref{eq:Xvec} and $F^{(\alpha)}$ is a function that performs the matrix multiplications of equation \ref{eq:usadelriccati} and extracts the appropriate element. Similarly, the boundary conditions become in the Riccati parametrization:
\begin{align}\label{eq:klriccati}
\vec{n}\cdot\nabla\g_i = \mp \frac{1}{L_i \zeta_i}(1 - \g_i \gt_j)N_j(\g_i - \g_j) + 2i\vec{n}\cdot\vec{A}\g_i
\end{align}
where the negative sign should be used for a boundary where region $j$ is to the right of region $i$, and the positive sign for a boundary where region $j$ is to the left of region $i$. A similar expression is found for $\vec{n}\cdot\nabla\gt_i$ by applying the $\tilde{\ldots}$ operation to equation \ref{eq:klriccati}. These are Neumann boundary conditions of the type

\begin{align}\label{eq:bc1reduced}
\vec{n}\cdot\nabla \chi^{(\alpha)} = B^{(\alpha)}(\g, \gt)
\end{align}
where $B^{(\alpha)}$ works in a similar manner as $F^{(\alpha)}$.

By multiplying equation \ref{eq:usadelreduced} by a test function $\eta(\vec{r})$ and integrating over the domain $\Omega$ in which the equations are defined, one gets what is called the weak formulation of the NLDEs (not to be confused with the weak proximity effect approximation):

\begin{align}\label{eq:weakform}
-\int_{\Omega} d\vec{r}\; \nabla \chi^{(\alpha)}\cdot\nabla\eta + \int_{\Omega} d\vec{r}\; F^{(\alpha)}(\g, \gt, \nabla\g, \nabla\gt)\eta + \int_{\partial\Omega}dS\; \vec{\nu}\cdot\nabla \chi^{(\alpha)}\eta = 0
\end{align}
where the divergence theorem has been used and $\partial\Omega$ is the boundary of $\Omega$. The unit vector $\vec{\nu}$ is an outward pointing surface normal, and is either parallel or antiparallel with the normal vector $\vec{n}$ as defined in the Kupriyanov-Lukichev boundary conditions. It may thus be expressed as $\vec{\nu} = (\vec{\nu}\cdot\vec{n})\vec{n}$.

 It is assumed that the domain $\Omega$ can be discretized into a mesh of $N_{el}$ elements, i.e., $N_{el}$ subdomains $\Omega_n$, so that equation \ref{eq:weakform} becomes
 

\begin{align}\label{eq:elweakform}
\sum_{n = 1}^{N_{el}}\int_{\Omega_n} d\vec{r}\;\left[- \nabla \chi^{(\alpha)}\cdot\nabla\eta + F^{(\alpha)}(\g, \gt, \nabla\g, \nabla\gt)\eta\right] + (\vec{\nu}\cdot\vec{n})\int_{\partial\Omega}dS\; \vec{n}\cdot\nabla \chi^{(\alpha)}\eta = 0
\end{align}

So far, no approximations have been made, and provided it is continuous in $\Omega_n$, the exact solution of equation \ref{eq:elweakform} exists in the infinite space of polynomials $P(\Omega_n)$. To progress further, we will use the Galerkin method, a common finite element formulation technique treated in most books on the subject, e.g\cite{cook_book_02}. The method consists of restricting the space in which solutions are sought, from $P(\Omega_n)$ to a finite dimensional space of polynomials $P^N(\Omega_n)$ consisting of all polynomials of degree $N$ or lower. Normally, $N$ is equal to 1 or 2.

On each element there are defined $N_n$ nodes, containing the degrees of freedom of the system - in this case the solution of the Usadel equation at the location of the node - and it is possible to define $N_n$ polynomials, $\phi_j(\vec{r})$, that interpolate between them. These interpolation functions span the space of $P^{N}(\Omega_n)$ and are used as a basis for the approximate solution of equation \ref{eq:elweakform}:

\begin{align}\label{eq:approx}
\chi^{(\alpha)} \approx X^{(\alpha)} = \sum_{j = 1}^{N_n} X^{(\alpha)}_j\phi_j
\end{align}
where $X^{(\alpha)}_j$ are the expansion coefficients for the approximate solution of equation $\alpha$. Furthermore, the test function $\eta$ is selected as

\begin{align}\label{eq:testf}
\eta = \sum_{j=1}^{N_n} \phi_j
\end{align}

We now consider the boundary term. Having meshed the domain $\Omega$, it is obvious that some of the element domains $\Omega_n$ intersect with the boundary $\partial \Omega$. In fact, the boundary is the union of all these intersections. It follows that the nodes associated with these intersections also lie on the boundary, and so there are defined interpolation functions also here. With the dimensionality of $\partial \Omega$ being one less than $\Omega$, the surface interpolation functions $\phi^S_j$, which are zero everywhere but on the boundary, are found by evaluating the element interpolation functions at the surface, i.e., $\phi^S_j = \phi_j(\vec{r}^S)$ where $\vec{r}^S$ is a surface coordinate.

With the approximation given in \ref{eq:approx}, equation \ref{eq:elweakform} is in general not satisfied, so that for every element the right hand side becomes equal to a residual, $R_j^{(\alpha)}$:

\begin{align}\label{eq:residual}
R_j^{(\alpha)} = 
\int_{\Omega_n} d\vec{r}\;\left[- \nabla X^{(\alpha)}\cdot\nabla\phi_j + F^{(\alpha)}(\g, \gt, \nabla\g, \nabla\gt)\phi_j \right] + (\vec{\nu}\cdot\vec{n})\int_{\partial\Omega}dS\; B^{(\alpha)}(\g, \gt)\phi^S_j
\end{align}

Equation \ref{eq:residual} is to be solved for $X^{(\alpha)}_j$ so that $R_j^{(\alpha)}=0$, however due to the nonlinearities introduced by $F^{(\alpha)}$ and $B^{(\alpha)}$ this needs to be done iteratively by Newton-Raphson iterations:

\begin{align}\label{eq:NRiter}
\left(X_i^{(\alpha)}\right)_{k+1} = \left(X_i^{(\alpha)}\right)_{k} - \left[J_{ij}^{(\alpha \beta)}\right]^{-1}\left(R_j^{(\beta)}\right)_k
\end{align}
with $J_{ij}$ the Jacobian matrix in the 8 dimensional parameter space, given as

\begin{align}
J_{ij}^{(\alpha \beta)} = \frac{\partial R_j^{(\beta)}}{\partial X_i^{(\alpha)}}
=\int_{\Omega_n} d\vec{r}\;\left[- \delta_{\alpha \beta}\nabla \phi_i\cdot\nabla\phi_j + \frac{\partial F^{(\beta)}}{\partial X^{(\alpha)}_i} \phi_i\phi_j \right] + (\vec{\nu}\cdot\vec{n})\int_{\partial\Omega}dS\; \frac{\partial B^{(\beta)}}{\partial X^{(\alpha)}_i} \phi^S_i\phi^S_j
\end{align}

Finally, \ref{eq:NRiter} needs to be assembled into a global system of equations by summing over all elements, taking element connectivity into account. This involves restructuring and expanding the element matrices into a global system matrix:

\begin{align}
\mathbb{X}_{k+1} = \mathbb{X}_k - \mathbb{J}^{-1}\mathbb{R}_k
\end{align}
where $\mathbb{J}$ is an $8M\times8M$ matrix, and $M$ is the number of nodes in the system. The integrals over the element domains are performed by changing coordinates to a reference element, and integrating numerically by means of a Gauss quadrature. This puts restrictions on how distorted a mesh can be, as the Jacobian for the coordinate transformation has to exist. In general, a structured mesh where the deviation from the geometry of the reference element is small will often give higher accuracy and reduce the computation time as the sparsity of the assembled matrices is increased.

\section*{Results}\label{sec:results}
\subsection*{Application: 2D and 3D superconductor/ferromagnet junctions}\label{sec:resultsapp}
The main advantage of the finite element method over the finite difference method, a method commonly used to solve partial differential equations numerically, is that it is formulated entirely without specifying element type, interpolation functions, spatial dimension or the geometry. This gives it the flexibility to solve PDEs on geometries which would be challenging to solve with the finite difference method. Here, we have used second order Lagrange polynomials as interpolation functions with quadrilateral (QUAD9) and hexagonal (HEX27) elements in 2D and 3D respectively. We illustrate this in the following. For the numerical implementation, we use the finite element library libMesh\cite{libmeshpaper} and its integration with the PETSc library of numerical equation solvers\cite{petsc-user-ref,petsc-efficient}. In the following, the superconducting regions will be treated as reservoirs such that the bulk expression for the Green function $\check{g} = \check{g}_\text{BCS}$ will be used. The superconductors thus effectively enter the problem as boundary conditions.

\subsubsection*{2D Josephson junction with external magnetic flux}
It is well known that for a Josephson junction where an external flux is applied to the intermediate region, the supercurrent exhibits a Fraunhofer interference pattern. In Refs. \cite{cuevas_prl_07, cuevas_jlt_08} a 2D superconductor/normal/superconductor Josephson junction was studied in the presence of an external magnetic flux. The authors revealed that the Fraunhofer interference pattern would qualitatively change its dependence on the external flux depending on the width of the junction $W$ relative to its length $L$. When $W \gg L$ a conventional Fraunhofer pattern was found, when $W \ll L$ the supercurrent was monotonically decaying. Moreover, it was shown that  the Fraunhofer interference pattern was accompanied by a regular array of proximity-induced vortices in the transversal direction of the normal metal region. The vortices are not present in the narrow width limit. Experimental verification of the appearance of proximity-induced vortices was recently reported in Ref. \cite{roditchev_np_2015} which considers Josephson junctions generated by a network of superconducting nanocrystals.

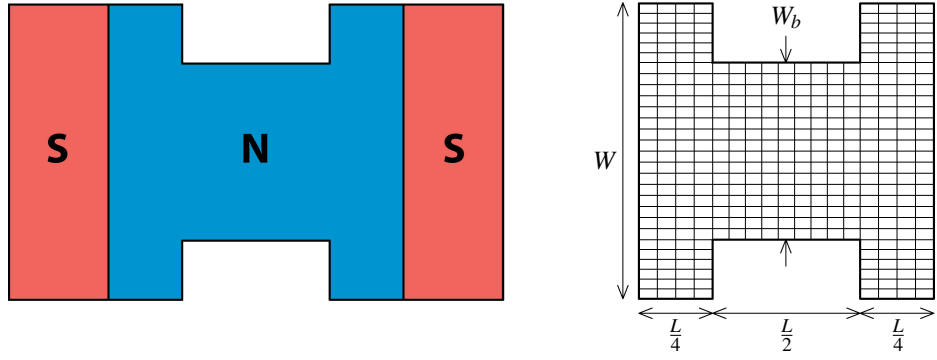
\begin{figure}
\centering
\begin{tikzpicture}
\node[anchor=south west,inner sep=0] (image) at (0,0) {\includegraphics[width=0.7\textwidth]{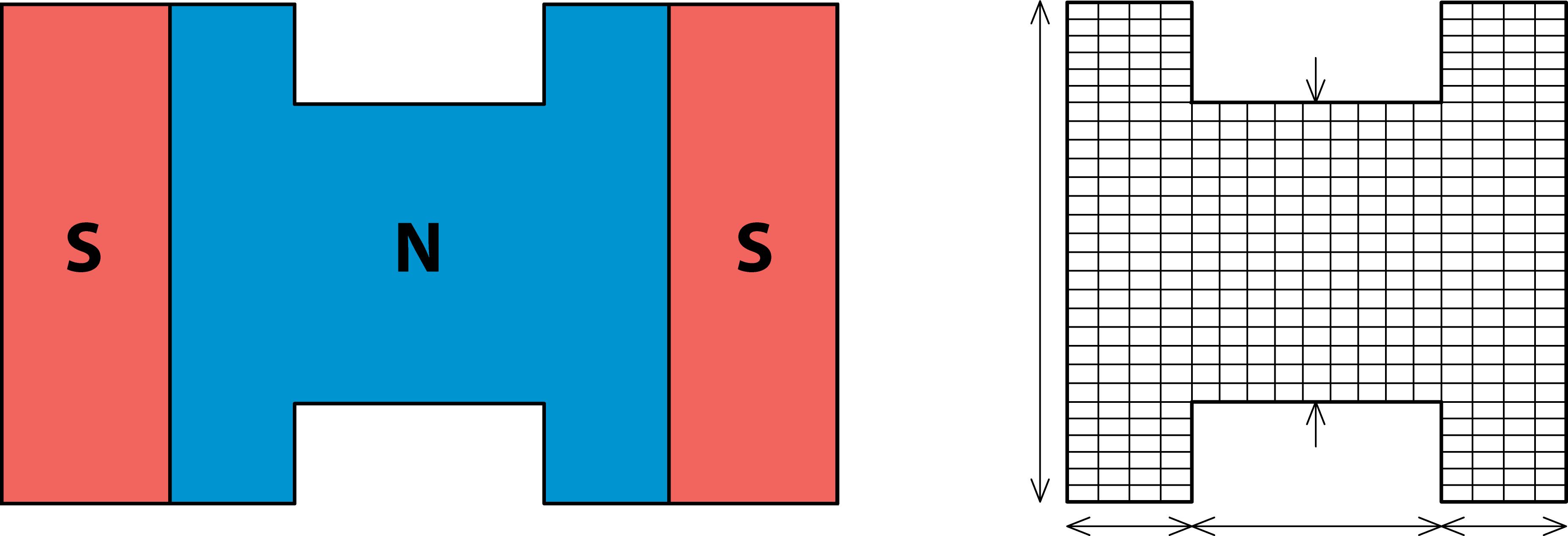}};
\begin{scope}[x={(image.south east)},y={(image.north west)}]
	\node at (0.84,0.95) {$W_b$};
	\node at (0.645,0.5) {$W$};
	\node at (0.72,-0.05) {$\frac{L}{4}$};
	\node at (0.84,-0.05) {$\frac{L}{2}$};
	\node at (0.96,-0.05) {$\frac{L}{4}$};
\end{scope}
\end{tikzpicture}
\vspace{0.2cm}
\caption{The geometry considered. To the left is shown a general outline of the Josephson junction, and to the right a typical mesh used in the numerical analysis.}
\label{fig:bottleneck}
\end{figure}

Here, we explore how the vortices disappear from the system as the width is reduced and the system transitions to a vortex-less state. We also determine how a change in the phase difference between the superconducting leads affect the vortices, and will show that this does not always correspond to a shift of the vortex array along the transverse direction.
 To illustrate the ease with which the finite element method handles non-trivial geometries, we consider a Josephson junction with a bottleneck in the normal metal region, as shown in Fig. \ref{fig:bottleneck}. We assume that the currents in the system are small, so that the magnetic field remains unaffected. As will be shown, it turns out to be possible to tune the geometry of the array along which the superconducting vortices align, swapping from a vertical necklace to a horizontal row of vortices and vice versa. Moreover, we demonstrate that changing superconducting phase difference, tunable \eg via a current-bias, causes vortices to merge. This offers an interesting route to exerting external control over topological excitations in superconducting hybrid structures.

\begin{figure}[b!]
\centering
\includegraphics[width=\textwidth]{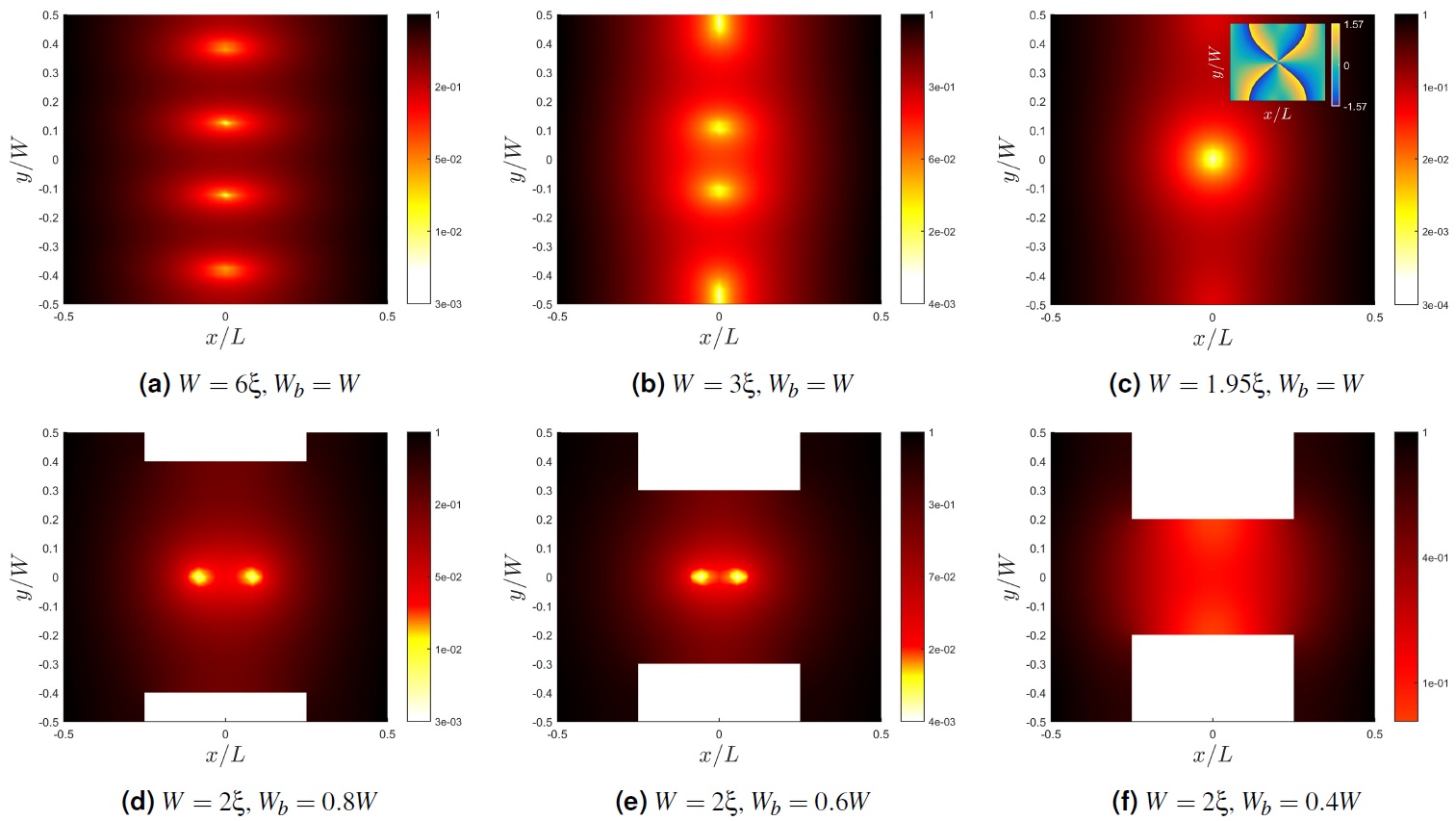}
\caption{The absolute value of the pair correlation function for different values of the width $W$ and bottleneck width $W_b$. The length is $L = 2\xi$. The inset of \textbf{2c} shows the phase of the pair correlation function. }
\label{fig:BNwidth}
\end{figure}

Fig. \ref{fig:BNwidth} shows the results for varying widths of the normal metal with an applied external flux of $\Phi = 4\Phi_0$, where $\Phi_0 = \frac{2e}{h}$ is the flux quantum. The flux is specified with respect to a rectangular cross section $W \times L$. The resistance ratio in the Kupriyanov-Lukichev boundary conditions is $\zeta = 3$ in both interfaces with the superconductors, and the temperature is $k_BT = 0.001\Delta$. All lengths are in units of the superconducting coherence length $\xi$.

It is seen that with no bottleneck, and with $W \gg L$, a linear array of vortices along the $y$-axis is found. This is shown in Fig. \ref{fig:BNwidth}\textbf{a} and is in agreement with Refs.\cite{cuevas_prl_07, cuevas_jlt_08} The number of vortices is simply equal to the number of flux quanta in the system. Furthermore, the fact that the vortices align themselves in an array implies that they repel, a feature they share with the Abrikosov vortices found in type II superconductors. Decreasing the width, pushes the vortices closer together which is energetically less favorable. With no phase difference between the superconducting leads, the phase correlation function is symmetric about both the $x$- and $y$-axis. This means that when the system becomes too narrow to sustain four vortices, two vortices must simultaneously translate vertically out of the system in a way which maintains this symmetry, as seen in Fig. \ref{fig:BNwidth}\textbf{b}. The two remaining vortices are seen to be forced closer together until they eventually meet at the origin, from which it may be inferred that for the given flux and geometry, the presence of two vortices is energetically favorable regardless of their separation. In particular, it is observed that within numerical precision, the vortices are found to completely overlap in Fig. \ref{fig:BNwidth}\textbf{c}, resulting in a single vortex. The winding of the phase of the pair correlation function along a contour around this vortex is found to be $4\pi$, implying a topological charge of 2, see inset of Fig. \ref{fig:BNwidth}\textbf{c}.   As the bottleneck is introduced, and the width further decreased, the vortices split symmetrically along the $x$-axis, as shown in Figs. \ref{fig:BNwidth}\textbf{d-e}. This behavior may also be explained by the symmetry of the system, which restrains the positions of the two vortices to be symmetric about the origin, on either the $x$-axis or the $y$-axis. As the vortices evidently feel a stronger repulsion from the edges than from each other, they are pushed together. However once they meet at the origin, they are free to separate along the $x$-axis and thus reduce the energy in the system. By continuing to decrease the bottleneck width, $W_b$, a point where even two vortices may not be sustained is eventually reached. The boundary conditions that constitute the superconducting leads enforce a constant pair correlation, and so it becomes increasingly difficult to maintain the curvature necessary for the vortices to exist as one approaches the superconductors. In other words, the vortices may not in a continuous fashion exit the system along the $x$-axis. Instead, the vortices are seen to return to the $y$-axis, and be expelled vertically.

While the vortices separate along the $x$-axis for decreasing bottleneck width, the length of the narrowing area is large enough to contain them, and so the system behaves as if the width is uniformly decreased. It has been verified that by reducing the horizontal extent of the bottleneck, it is possible to create a situation where the vortices are pushed to the wide regions of the junction, at which point they become virtually independent of the bottleneck width $W_b$.

\begin{figure}[H]
\centering
\includegraphics[width=\textwidth]{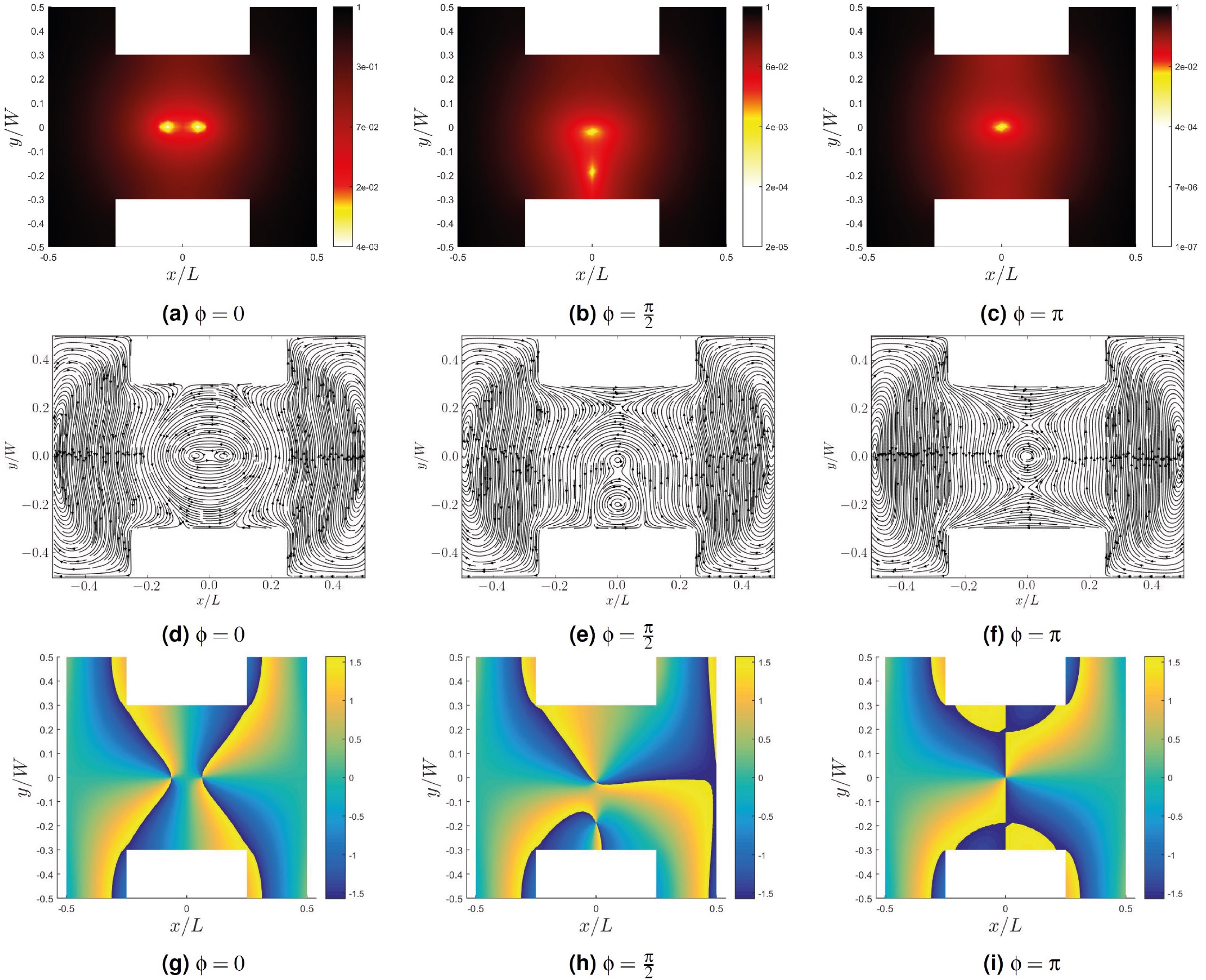}
\caption{Results based on the solution of the Usadel equation for varying phase difference between the superconductors, with $L=W=2\xi$, $W_b = 0.6W$ and $\Phi=4\Phi_0$. \textbf{(a)}-\textbf{(c)} The absolute value of the pair correlation function, \textbf{(d)}-\textbf{(f)} the current density, \textbf{(g)}-\textbf{(i)} the phase of the pair correlation function.}
\label{fig:BNPC}
\end{figure}

We also show the results for varying phase difference between the superconductors, for a geometry with $W_b = 0.6W$, shown in Fig. \ref{fig:BNPC}. We find that not only are the positions of the vortices changed by varying the phase difference $\phi$, but also the number of vortices is altered. Figs. \ref{fig:BNPC}\textbf{a-c} show the absolute value of the pair correlation function. With no phase difference, two vortices are located symmetrically along the $x$-axis. As $\phi$ increases from 0, the two vortices coalesce at the origin. Further increase translates one of the vortices in the negative $y$-direction, until only a single vortex remains. We have confirmed that the spatial rearrangement of the vortex pattern and the merging of vortices also takes place even without the bottleneck geometry, i.e. for a rectangular N region. The dependence of the vortex positions on the phase difference may be explained by the magnetic field, which in the small current approximation, permeates the normal metal unhindered. The current generated by the phase difference is altered by the field which in turn influences which locations that are energetically favorable for the vortices.

The current density for each of the cases considered are shown in Figs. \ref{fig:BNPC}\textbf{d-f}. Close to a vortex, where pair correlation is low, currents are induced by the magnetic field and circulate counter-clockwise. Due to the omnipresent magnetic field, screening currents circulating clockwise are generated which dominate when pair correlation is high. As the pair correlation function is weakened upon approaching the vortex core, one observes an abrupt change in the current density pattern at a certain distance from the vortex.

The phase of the pair correlation function is given as $\theta = \arctan(\Im{\Psi} / \Re{\Psi})$.
By integrating $\nabla \theta$ along a contour going around a point where the pair correlation function vanishes, a value of $2\pi$ is found. This can be seen directly from the phase plots in Figs. \ref{fig:BNPC}\textbf{g-i}, as any curve around a zero of the pair correlation function has to traverse two discontinuous jumps of value $\pi$. In other words, these points have a topological charge of one, showing that they are indeed vortices. With the approximation of weak currents we do not however find flux quantization, as this requires a self consistent calculation of the magnetic field. 


\begin{figure}
\centering
\includegraphics[width=0.8\textwidth]{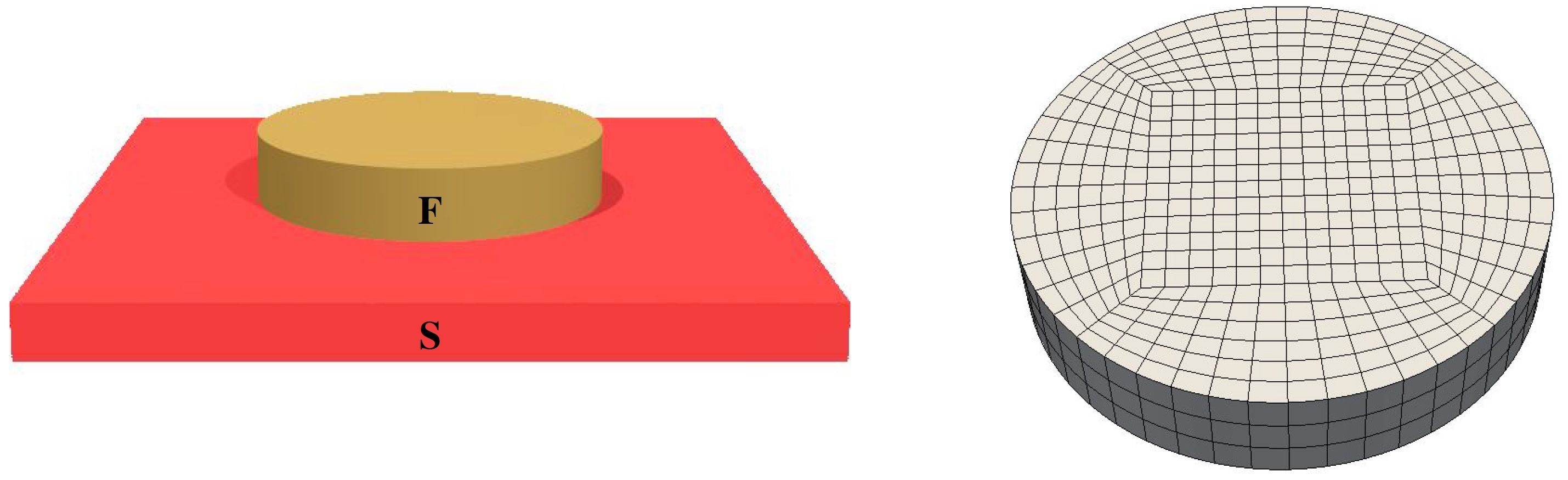}
\caption{To the left is shown the 3D geometry considered. The superconductor is assumed to have an infinite extent and is included only as a boundary condition. The mesh used when solving the Usadel equation in the ferromagnet is shown to the right. The radius is $R = 2\xi$, and the thickness is $L_z = 0.4R$. }
\label{fig:3dmodel}
\end{figure}

\newpage
\subsubsection*{3D ferromagnetic nanoisland}
We also demonstrate how the finite element method is capable of dealing with fully three-dimensional structures with non-rectangular geometry by considering a superconductor/ferromagnet bilayer as depicted in Fig. \ref{fig:3dmodel}. The ferromagnet is cylindrical with a radius of $R = 2\xi$ and a height of $L_z = 0.4R$, and is placed atop an assumed infinite superconductor. Such a geometry is inspired by Ref. \cite{lange_prl_03} which experimentally explores the appearance of magnetic field induced superconductivity in a lattice of ferromagnetic islands placed on top of a superconductor. While the experimental setup is far too sophisticated for their results to be recreated by the example considered herein, it does demonstrate the relevance of the model.

\begin{figure}[H]
\centering
\includegraphics[width=0.9\textwidth]{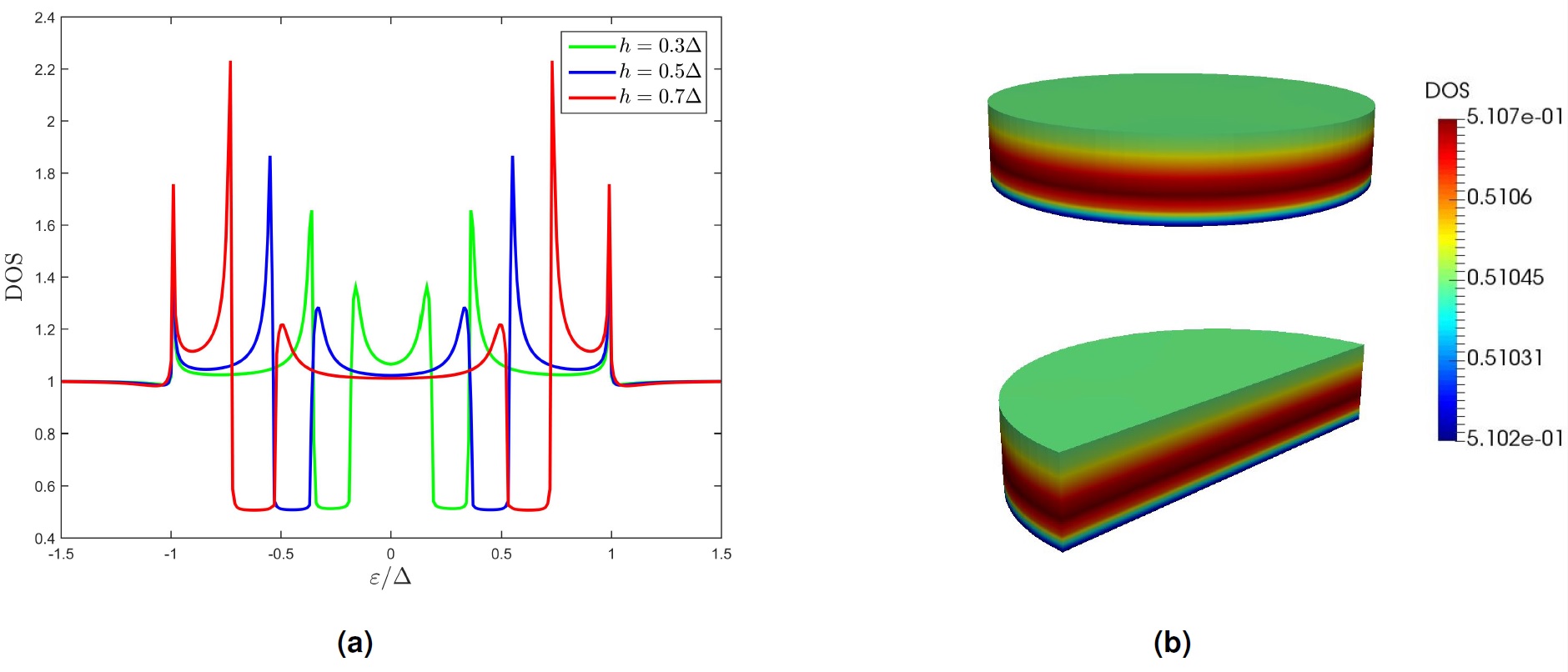}
\caption{\textbf{a} Density of states for the 3D ferromagnet structure for various strengths of the vertical exchange field, \textbf{b} spatial distribution of the density of states for energy $\varepsilon = 0.5\Delta$ and exchange field $h=0.5\Delta$.}
\label{fig:3D_DOS}
\end{figure}

We use Kupriyanov-Lukichev boundary conditions with a resistance ratio of $\zeta = 3$. We compute the density of states (DOS) for this structure with an exchange field $h$ equal to 0.3$\Delta$, 0.5$\Delta$ and 0.7$\Delta$ in the vertical direction, as shown in Fig. \ref{fig:3D_DOS}\textbf{a}. The results are identical with the one-dimensional solution to the S/F bilayer, displaying an enhanced DOS at the Fermi level $(\varepsilon=0)$ and a spin-split minigap structure \cite{buzdin_prb_00, zareyan_prb_02, kontos_prl_01, yokoyama_prb_07, linder_prb_08, sangiorgio_prl_08, linder_scirep_15}. The spatial distribution of the DOS is nearly constant for this particular parameter set choice in F as illustrated by figure \ref{fig:3D_DOS}\textbf{b}, thus proving the correctness of the method.

\subsubsection*{3D ferromagnet with superconducting islands}
\begin{figure}[H]
\centering
\includegraphics[width=\textwidth]{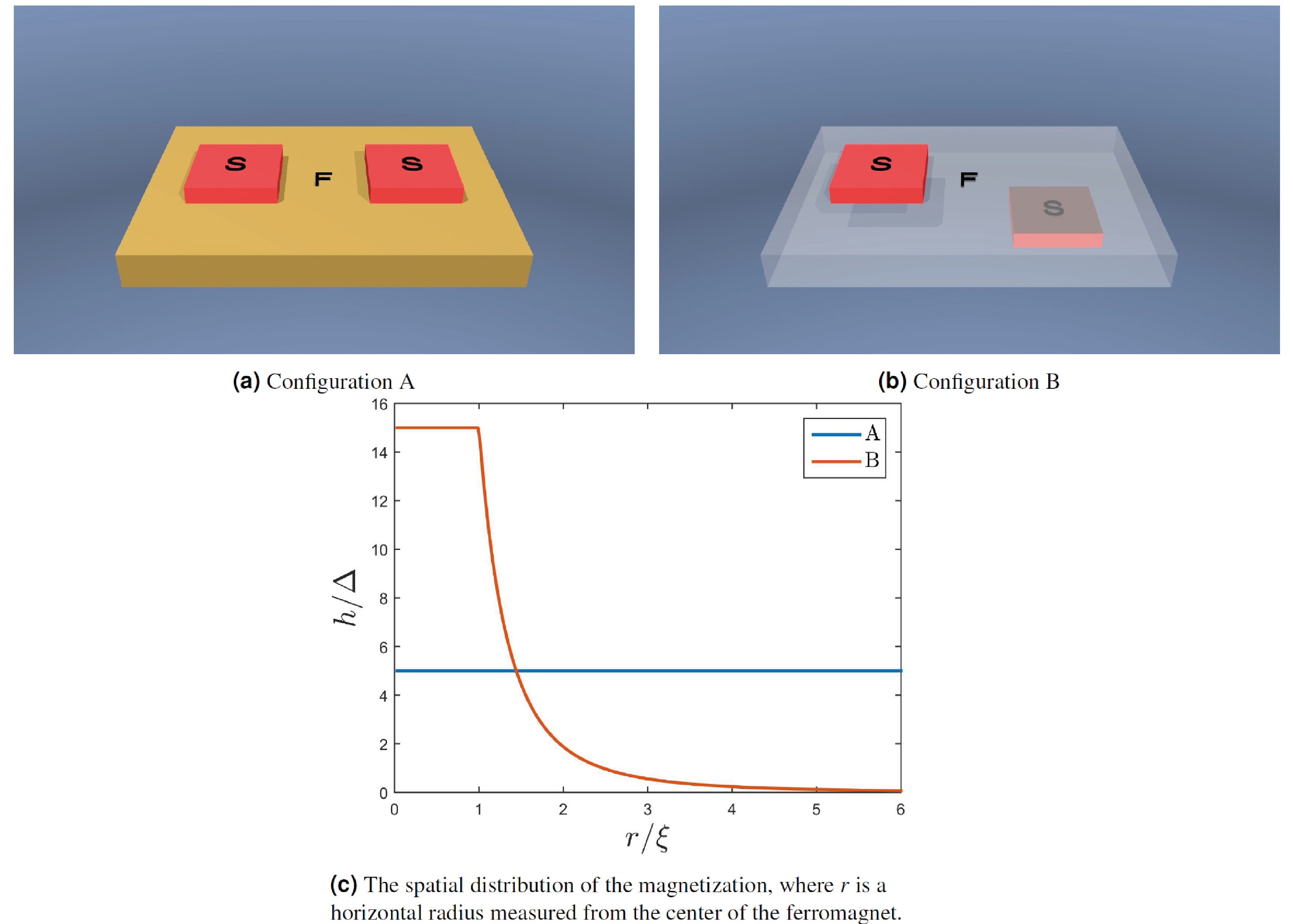}
\caption{The geometries and magnetization considered. Marked in red are superconducting islands placed on a ferromagnet. The dimensions of the ferromagnet is $L_x\times L_y \times L_z = 10\xi \times 7\xi \times \xi$. In subfigure \textbf{b} the ferromagnet has been made transparent for visualization purposes.}
\label{fig:SFSgeometry}
\end{figure}

To illustrate the 3D capabilities of the method developed on a system which cannot be described by an effective 1D model, we consider two variations of a system where two superconducting islands are placed on a ferromagnet with dimensions $L_x \times L_y \times L_z = 10\xi \times 7\xi \times \xi$, as shown in Fig. \ref{fig:SFSgeometry}. To avoid self-consistency iterations, the islands are approximated by bulk BCS superconductors, and are included as Kupriyanov-Lukichev boundary conditions with a resistance ratio of $\zeta = 1.5$.  The dimensions of the islands are $2.5\xi \times 2.5\xi$. The motivation for these analyses is to study the current flow between the superconducting islands and the spatial modulation of the density of states due the proximity effect in the presence of a supercurrent. To this end, the islands are given a phase difference of $\phi = \frac{\pi}{2}$. The configurations considered are:
\begin{itemize}
\item[A] The superconducting islands are placed on the same side of the ferromagnet with a separation of $2.5\xi$, as shown if Fig. \ref{fig:SFSgeometry}\textbf{a}. The ferromagnet has a constant magnetization of $h = 5\Delta$ in the vertical direction.
\item[B] One of the superconducting islands is moved to the opposite side of the ferromagnet, shown in Fig. \ref{fig:SFSgeometry}\textbf{b}. The magnetization is pointing in the vertical direction, with a spatial distribution shown in Fig. \ref{fig:SFSgeometry}\textbf{c}.  
\end{itemize}

Configuration A may be realized experimentally by growing the ferromagnetic film on a substrate and placing superconducting electrodes on top of it. Configuration B may be created by placing the lower superconducting electrode on the substrate and subsequently grow a normal metal film on top of it (a similar type of geometry was considered in the context of Fraunhofer patterns in Ref. \cite{alidoust_prb_13}). The upper superconducting electrode is then placed atop the film. The spatial distribution of the magnetization used in Configuration B can be generated by placing a strong ferromagnet on top of the normal film. This will magnetize the film across the thickness, creating a cross section approximately equal to the ferromagnet within which the magnetization is constant. The ability of the ferromagnet to induce magnetization within the normal film abates quickly as the distance from it increases, thus generating the distribution shown in Fig. \ref{fig:SFSgeometry}\textbf{c}.

\begin{figure}[H]
\centering
\includegraphics[width=0.9\textwidth]{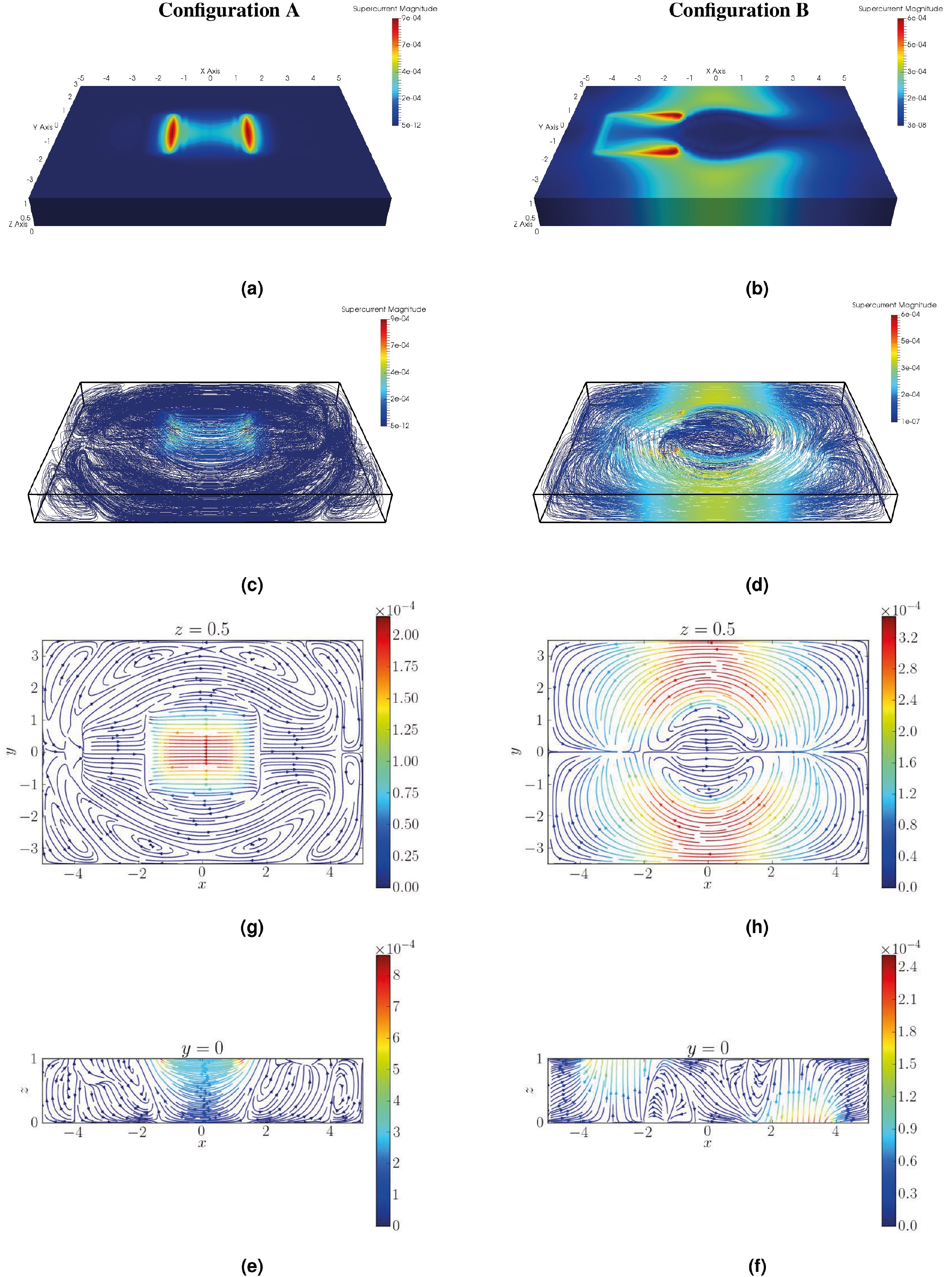}
\caption{The supercurrent present in both Configuration A and Configuration B. All currents are scaled by $J_0 = \frac{N_0eD\Delta}{8}$ and all lengths by $\xi$.}
\label{fig:SFSResults}
\end{figure}

The results from both configurations are given in Fig. \ref{fig:SFSResults}. For Configuration A it is seen that the current is largely confined to the region between the islands, passing from one to the other, shown in Figs. \ref{fig:SFSResults}\textbf{a} and \ref{fig:SFSResults}\textbf{g}. Due to the magnetization being uniform, the supercurrent travels along a path that minimizes the distance, and thus has no component pointing along the transversal direction (the $y$-axis of Fig. \ref{fig:SFSResults}). However, as the current enters and exits the ferromagnet vertically, it is seen to arc into the thickness of the film, as seen in Figs. \ref{fig:SFSResults}\textbf{c} and \ref{fig:SFSResults}\textbf{e}. 

Configuration B has a magnetization of $h(\vec{r}) = 15\Delta$ within a horizontal radius $\vec{r}$ of one superconducting coherence length $\xi$ from the center of the ferromagnet. This has a significant effect on the supercurrent. The supercurrent flows vertically into the system from the lower superconductor, as seen in Fig. \ref{fig:SFSResults}\textbf{d}. However, rather than flowing directly to the upper superconductor, as would have been the case for a homogeneous magnetization, the current is seen to avoid the area of highest magnetization by following a semicircular path, shown in Figs. \ref{fig:SFSResults}\textbf{b}, \ref{fig:SFSResults}\textbf{d} and \ref{fig:SFSResults}\textbf{h}. The exchange field has a detrimental effect on the superconducting correlations as it breaks up the Cooper pairs. For this reason it is natural that the path selected by the supercurrent eventually transitions from the shortest route, to a path where the central area is avoided as $h$ increases. In this sense, the exchange field is seen to influence the supercurrent in a way which is analogous to the way a resistance influences a normal current. It is interesting that this transition has occurred already for $h = 15\Delta$, which is to be considered a somewhat weak magnetization, and may provide means for customizing the supercurrent path.

\begin{figure}[H]
\centering
\includegraphics[width=\textwidth]{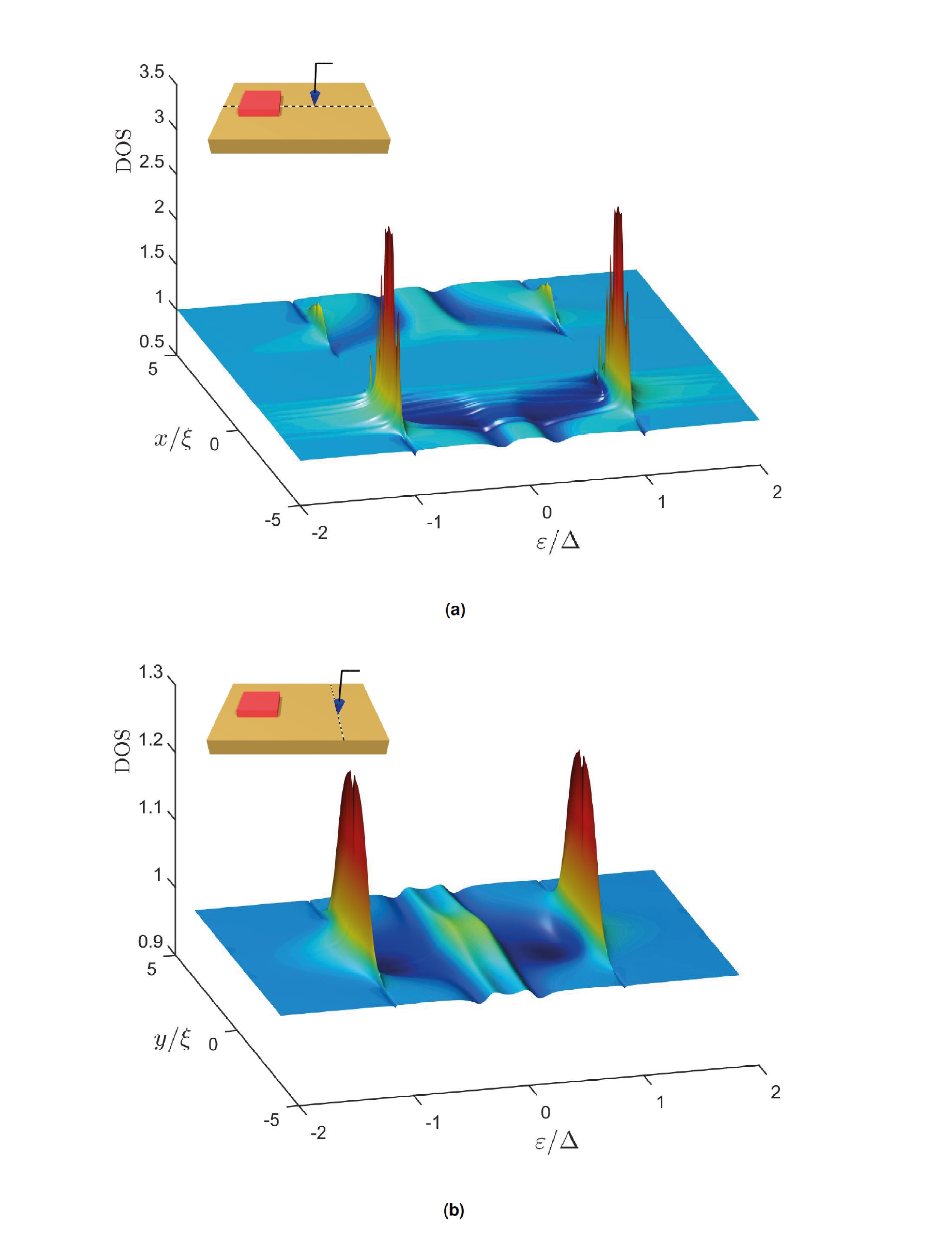}
\caption{The density of states along two different lines on the surface of the ferromagnet.}
\label{fig:SFSDOS}
\end{figure}

In Fig. \ref{fig:SFSDOS} we show the density of states (DOS) along two different lines on the surface of the ferromagnet in Configuration B, thus simulating the measurement of $\frac{\text{d} I}{\text{d} V} \propto \text{DOS}$ by scanning tunneling microscopy. The DOS is seen to feature a strong spatial modulation. Along the $x$-axis, as seen in Fig. \ref{fig:SFSDOS}\textbf{a}, the probed line passes directly underneath the upper superconductor. Here the characteristic peaks associated with the superconducting DOS are observed at $\varepsilon = \pm \Delta$. Similar peaks are also created on the surface above the lower superconductor. Furthermore, a slight suppression of the DOS is found in the same regions, at the level of $\varepsilon = \pm h(\vec{r})$, which is typical for superconductor/ferromagnet hybrid structures \cite{linder_prb_08}. The second line is placed opposite the lower superconductor, in the $y$-direction as shown in Fig. \ref{fig:SFSDOS}\textbf{b}. Also here, the characteristic peaks and split gap is found. The proximity effect is seen to decay as one moves away from the position of the superconductor, so that the DOS approaches that of a normal metal, which is reasonable.

\section*{Conclusion}\label{sec:discussion}
We have demonstrated how the full, spin dependent, Usadel equation may be solved by the finite element method. The method excels in solving differential equations for non-trivial geometries and may find use in solving a wide range of problems which have not been manageable with other methods.
A natural development of the finite element method presented herein would be to incorporate the kinetic equations coming from the Keldysh part of the quasiclassical equations in non-equilibrium situations. The methodology may also be generalized to handle time dependent problems such as domain wall motion. Work is currently ongoing on these subjects which may find interesting applications in the field of superconducting spintronics \cite{linder_nphys_15, eschrig_physrep_15}.

\newpage

\section*{Acknowledgements}

We thank J. A. Ouassou for useful discussions. J.L was supported
by the Research Council of Norway, Grants No. 205591, 216700, 240806 and the "Outstanding Academic Fellows" programme at NTNU. 

\section*{Author contributions statement}
M.A. developed the numerical code and performed the calculations. Both contributed to the discussion of the results and the writing of the manuscript.

\section*{Additional information}

\textbf{Competing financial interests} The authors declare no competing financial interests.

\end{document}